\documentclass[12pt]{iopart}

\usepackage{color}

\begin{document}

\title{Information content versus word length in random typing}

\author{Ramon Ferrer-i-Cancho$^{1,*}$ and Ferm\'{i}n Moscoso del Prado Mart\'{i}n$^{2,3,4}$} 

\address{$^1$ Complexity \& Quantitative Linguistics Lab \\
Departament de Llenguatges i Sistemes Inform\`atics, \\
TALP Research Center, Universitat Polit\`ecnica de Catalunya, \\
Campus Nord, Edifici Omega Jordi Girona Salgado 1-3. \\
08034 Barcelona, Catalonia (Spain)}

\address{$^2$ Laboratoire de Psychologie Cognitive (UMR6146) \\
CNRS \& Aix-Marseille Universit\'{e} I, Marseille, France}
\address{$^3$ Laboratoire Dynamique du Langage (UMR5596) \\
CNRS \& Universit\'{e} de Lyon II, Lyon, France}
\address{$^4$ Institut Rh\^{o}ne-Alpin de Syst\`{e}mes Complexes, Lyon, France}

\eads{\mailto{rferrericancho@lsi.upc.edu} and \mailto{fermin.moscoso-del-prado@gmail.com}}

\begin{abstract}
Recently, it has been claimed that a linear relationship between a measure of information content and word length is expected from word length optimization and it has been shown that this linearity is supported by a strong correlation between information content and word length in many languages
({\em Piantadosi et al. 2011, PNAS 108, 3825-3826}). Here, we study in detail some connections between this measure and standard information theory. The relationship between the measure and word length is studied for the popular random typing process where a text is constructed by pressing keys at random from a keyboard containing letters and a space behaving as a word delimiter. Although this random process does not optimize word lengths according to information content, it exhibits a linear relationship between information content and word length. The exact slope and intercept are presented for three major variants of the random typing process. A strong correlation between information content and word length can simply arise from the units making a word (e.g., letters) and not necessarily from the interplay between a word and its context as proposed by Piantadosi {\em et al}. In itself, the linear relation does not entail the results of any optimization process. 
\end{abstract}



\submitto{Journal of Statistical Mechanics: theory and experiment (JSTAT)}

\maketitle

\section{Introduction}

In his pioneering research, G. K. Zipf showed that more frequent words tend to be shorter \cite{Zipf1935}, and parallels of this brevity law have been reported for the behavior of other species \cite{Semple2010a, Ferrer2009g}. Recently, it has been argued that "average information content is a much better predictor of word length than frequency" and that this "indicates that human lexicons are efficiently structured for communication by taking into account interword statistical dependencies." \cite[p.~1]{Piantadosi2011a}.
According to the uniform information density hypothesis (e.g., \cite{Jaeger2010a}), "language users
make choices that keep the number of bits of information communicated 
per unit of time approximately constant" and thus "the amount of information conveyed by a word should be linearly related to the amount of time it takes to produce --approximately, its length-- to convey the same amount of information
in each unit of time" \cite[p.~1]{Piantadosi2011a}.
Here it will be shown that hitting keys from a keyboard at random (e.g., \cite{Miller1963, Li1992b}) generates words that reproduce this linear relationship. 
Therefore, the observation of such a linear relationship does not constitute unequivocal evidence for any kind of optimal choices made by speakers.

Throughout this paper, $C$ denotes contexts and $W$ denotes words. As in Ref.~\cite{Piantadosi2011a}, the context of a word consists of a fixed number of preceding words, and the information content of a word $w$ is given by
\begin{eqnarray*}
I(w) = - \sum_{c} p(C = c | W = w) \ln p(W = w | C = c).\label{information_content_equation}
\end{eqnarray*}
The expected information content of words of length $\ell$ is defined as \cite{Piantadosi2011a} 
\begin{eqnarray}
I(\ell)   & = & \sum_{\|w\|=\ell} p(W = w | \| w \| = \ell) I(w), \label{information_content_versus_length_equation}
\end{eqnarray}
where $\| w \|$ is the length (in letters) of a word $w$ and $\ell$ is a fixed parameter value. 
In this study, we detail some connections between $I(w)$ and standard information theory measures. The definition of $I(w)$ that we borrow from Ref.~\cite{Piantadosi2011a} is somewhat idiosyncratic in relation to standard information-theory. We found that, Ref.~\cite{Cohen2008a}, the reference supplied in Ref.~\cite{Piantadosi2011a} as a justification for Eq.~\ref{information_content_versus_length_equation}, does not in fact justify the equation in any evident way.
In this study we demonstrate that
$I(\ell)$ is a linear function of $\ell$ for a general class of random typing processes. The only requirement is that the context is defined by means of neighbouring words (as in \cite{Piantadosi2011a}) or that empty words (words of length zero) are allowed as in many variants of the random typing process \cite{Miller1963, Suzuki2004a, Conrad2004a}.

\section{Connections with standard information theory}

We now introduce our basic notation and conventions. The self-information of an event that has probability $p$ is $-\ln p$. We consider $C$ and $W$ independent if and only if $p(C = c , W = w) = p(C = c) p(W = w)$. As usual, by the definition of conditional probability, independence implies both $p(C = c | W = w) = p(C = c)$ and $p(W = w | C = c) = p(W = w)$, for any individual $c$ and $w$. Therefore, under independence between $C$ and $W$, it holds that $I(w) = I_0(w) = - \ln p(W = w)$, that is to say, $I(w)$ is just the self-information of $w$. The expected self-information content of a word of length $\ell$ is 
\begin{eqnarray}
I_0(\ell)  & = & - \sum_{\|w\|=\ell} p(W = w | \|w\| = \ell) \ln p(W = w) \nonumber \\
        & = & - \sum_{\|w\|=\ell} p(W = w | \|w\| = \ell) \ln p(W = w , \|w\| = \ell).
\label{expected_information_content_under_independence_equation}
\end{eqnarray}
In sum, under independence between $C$ and $W$, $I(\ell)$ and $I_0(\ell)$ coincide.

The conditional entropy is defined as,
\begin{eqnarray}
H(W|C) & = & \sum_{c} p(C = c) H(W| C = c) \nonumber \\
       & = & - \sum_{c} p(C = c) \sum_{w} p(W = w | C = c) \ln p(W = w | C = c). \label{word_conditional_entropy_equation}
\end{eqnarray}
Given only the joint probability, i.e. $p(W = w, C = c)$, one can use Bayes' Theorem for calculating the conditional and marginal probabilities, as it was done in previous work \cite{Piantadosi2011a} and is assumed by various information theoretic models of Zipf's law for word frequencies \cite{Prokopenko2010a, Ferrer2004e}. 
Simple application of Bayes' Theorem to the definition of $H(W|C)$ in \eref{word_conditional_entropy_equation} shows that the conditional entropy is the expectation of $I(w)$: 
\begin{eqnarray}
H(W|C) & = & - \sum_{c} \sum_{w} p(W = w , C = c) \ln p(W = w | C = c) \nonumber \\
       & = & - \sum_{w} p(W = w) \sum_{c} \frac{p(W = w , C = c)}{p(W =w)} \ln p(W = w | C = c) \nonumber \\
       & = & - \sum_{w} p(W = w) \sum_{c} p(C = c | W = w) \ln p(W = w | C = c) \nonumber \\
       & = & \sum_{w} p(W = w) I(w) = E[I(w)]. \label{relationship_between_word_conditional_entropy_and_information_content_equation}
\end{eqnarray}

It is not difficult to see that $I_0(w)$ is the upper bound of $I(w)$ and $H(C|w)$ is its lower bound; formally, 
\begin{eqnarray}
H(C|w) \leq I(w) \leq I_0(w).
\label{information_content_bounds_equation}
\end{eqnarray}
As for a lower bound of $I(w)$, the relative entropy (or Kullback-Leibler divergence) between the context conditional probability and the word conditional probability is \cite{Cover2006a}
\begin{eqnarray*}
\fl D(p(C = c | W = w) \| p(W = w | C = c)) & = & \sum_{c} p(C = c | W = w) \ln \frac{p(C = c | W = w)}{p(W = w | C = c)} \\
                                & = & \sum_{c} p(C = c | W = w) \ln p(C = c | W = w) \\
                                &   & - \sum_{c} p(C = c | W = w) \ln p(W = w | C = c) \\
                                & = & I(w) - H(C | w).
\end{eqnarray*} 
Therefore $I(w) \geq H(C | w)$ by the non-negativity of the relative entropy \cite{Cover2006a}.
As for the upper bound of $I(w)$, 
the non-negativity of mutual information, i.e. $I(W; C) = H(W) - H(W|C) \geq 0$ \cite{Cover2006a} and \eref{relationship_between_word_conditional_entropy_and_information_content_equation}, yields
\begin{eqnarray*}
H(W | C) & \leq & H(W) \\
\sum_{w} p(W=w)I(w) & \leq  & - \sum_{w} p(W=w) \ln p(W = w) \\
                    & =     & \sum_{w} p(W=w) I_0(w)    
\end{eqnarray*}
if and only if $I(w) \leq I_0(w)$, as we wanted to prove.
Combining \eref{information_content_versus_length_equation} and \eref{information_content_bounds_equation} results in 
\begin{equation}
I_C(\ell) \leq I(\ell) \leq I_0(\ell), 
\label{information_content_versus_length_bounds_equation}
\end{equation}
where $I_C(\ell)$ is defined as 
\begin{eqnarray*}
I_C(\ell)  & = & \sum_{\|w\|=\ell} p(W = w | \|w\| = \ell) H(C|w).
\end{eqnarray*}

\section{Information content versus length in random typing}

Random typing \cite{Miller1963, Conrad2004a} is a process in which a sequence of characters is produced by sampling randomly from a set of possible characters. Here we consider a generalized random typing model based upon variants allowing for unequal letter probabilities as in \cite{Li1992b,Conrad2004a} and allowing one to specify a minimum word length \cite{Ferrer2009a}.

Assume that characters are produced from an alphabet $\Sigma = \{\sigma_0, ..., \sigma_i, ..., \sigma_{\lambda-1}\}$, where $\lambda$ is the alphabet size, $\sigma_0$ represents the word delimiter (i.e., the space character) and the remaining characters of $\Sigma$ are letters. We assume that all the characters in $\Sigma$ are produced at random and independently, with the only exception that two instances of the space character must be separated by at least $\ell_0$ intervening characters other than the space. In such model, the production of a word is separated into two phases: generation of the space-free prefix of length $\ell_0$, and generation of the remainder. $S$ is a random variable taking values from $\Sigma$ as generated by the random typing process. $p_\Sigma(S = s)$ is defined as the probability of producing character $s$ as the $k$-$th$ character after the last space produced (or after the beginning of the sequence if no space has been produced yet), for any value $k \geq \ell_0$. $p_{\Sigma \setminus \{\sigma_0 \}}(S = s)$ is the same probability as $p_\Sigma(S = s)$ for values of $k < \ell_0$. The abbreviation $p_0 = p_\Sigma(S = \sigma_0)$ will be used hereafter.
We assume that $p_\Sigma(S = s) > 0$ for all characters in $\Sigma$ with the additional constraint that $p_0 < 1$. $p_{\Sigma \setminus \{\sigma_0 \}}(S = s)$ is defined in terms of $p_{\Sigma}(S = s)$,
\begin{eqnarray*}
p_{\Sigma \setminus \{\sigma_0 \}}(S = s) = 
   \left\{ 
       \begin{array}{cc}
           \frac{p_\Sigma(S = s)}{1 - p_0} & \qquad \mbox{if~} s \neq \sigma_0 \\
           $0$ &  \qquad \mbox{if~} s = \sigma_0.
       \end{array}
   \right.
\end{eqnarray*}
The generalized random typing process with unequal letter probabilities is defined by $\lambda$ parameters: $\ell_0$ and the $\lambda - 1$ probabilities  $p_\Sigma(S = \sigma_i)$ for $0 \leq i \leq \lambda - 2$ with 
\begin{equation*}
p_\Sigma(S = \sigma_{\lambda - 1}) = 1 - \sum_{i = 0}^{\lambda - 2} p_\Sigma(S = \sigma_i).
\end{equation*}
Notice the additional parameter $\ell_0$ that is not considered in other versions of the random typing model and allows for unequal character probabilities \cite{Li1992b,Conrad2004a}.

In the remainder of this section we start by proving that $I_0(\ell)$ is a linear function of $\ell$, providing exact analytical expressions for its slope and intercept. We continue by showing that $I(\ell)$ can be inferred from $I_0(\ell)$. If the context is defined by words, as in Ref.~\cite{Piantadosi2011a}, then $I(\ell) = I_0(\ell)$ because our generalized random typing process produces words independently from the previous ones. If the context are characters, then $I(\ell) = I_0(\ell)$ is also warranted when $\ell_0 = 0$ because this is the case where self-repulsion of the space is suppressed. When $\ell_0 > 0$, \eref{information_content_versus_length_bounds_equation} indicates that $I(\ell)$ cannot exceed $I_0(\ell)$.

In order to calculate the probability of producing a concrete word $w = s_1, ..., s_i, ..., s_\ell$, where $s_i$ is the $i$-th character from $\Sigma$ of $w$, we use the shorthand
\begin{eqnarray*}
{\cal P}_{i,j} = \prod_{h = i}^{j} p_\Sigma(S = s_h).
\end{eqnarray*}
By the independence between characters (except for space self-repulsion at distances smaller than $\ell_0$), the probability that a random word $W$ that has length $\ell$ coincides with $w = s_1, ..., s_i, ..., s_\ell$ is
\begin{eqnarray}
p(W = w, \|w\| = \ell) & =  & \left( \prod_{i=1}^{\ell_0} p_{\Sigma \setminus \{\sigma_0 \}}(S = s_i) \right) \left( \prod_{i = \ell_0 + 1}^\ell p_\Sigma(S = s_i) \right) p_0 \nonumber \\
                    & =  & \frac{p_0}{(1-p_0)^{\ell_0}} \left( \prod_{i= 1}^\ell p_\Sigma(S = s_i) \right) \nonumber \\
                    & =  & \frac{p_0}{(1-p_0)^{\ell_0}} {\cal P}_{1,\ell} , \label{extended_word_probability_equation}
\end{eqnarray}
the probability that a word has length $\ell$ is 
\begin{eqnarray*}
p(\|w\| = \ell) = p_0 (1-p_0)^{\ell - \ell_0}
\end{eqnarray*}
and the probability of a word $w$ given its length is therefore
\begin{eqnarray}
p(W = w | \|w \| = \ell) & = & \frac{p(W = w , \|w \| = \ell)}{p(\|w \| = \ell)} \nonumber \\
                          & = & \frac{1}{(1-p_0)^{\ell}} {\cal P}_{1,\ell}.
\label{extended_conditional_word_probability_equation}
\end{eqnarray}
Applying \eref{extended_word_probability_equation}, the self-information of a word $w$ of length $\ell$ is
\begin{eqnarray}
- \ln p(W = w, \|w\| = \ell) = b - \sum_{i=1}^\ell \ln p_{\Sigma}(S = s_i),
\label{extended_self_information_equation}
\end{eqnarray}
where $b$ is defined as 
\begin{eqnarray}
b = \ln \frac{(1 - p_0)^{\ell_0}}{p_0}.
\label{intercept_equation}
\end{eqnarray}
Combining \eref{extended_conditional_word_probability_equation} and \eref{extended_self_information_equation}
with the definition of $I_0(\ell)$ in \eref{expected_information_content_under_independence_equation}, gives
\begin{eqnarray*}
I_0(\ell) & = & \frac{1}{(1-p_0)^\ell} \sum_{s_1, ..., s_\ell} {\cal P}_{1,\ell} \left(b - \sum_{i=1}^\ell \ln p_\Sigma(S = s_i) \right).
\end{eqnarray*}
Bearing in mind that
\begin{eqnarray*}
\sum_{s_1, ..., s_\ell} {\cal P}_{1,\ell} & = & \sum_{s_1 \in \Sigma \setminus \{\sigma_0\}} ... \sum_{s_i \in \Sigma \setminus \{\sigma_0\}} ... \sum_{s_\ell \in \Sigma \setminus \{\sigma_0\}} {\cal P}_{1,\ell} \\
                                    & = & \sum_{s_1 \in \Sigma \setminus \{\sigma_0\}} ... \sum_{s_i \in \Sigma \setminus \{\sigma_0\}} ... \sum_{s_\ell \in \Sigma \setminus \{\sigma_0\}} \prod_{h = 1}^{\ell} p_\Sigma(S = s_h) \\
                                    & = & \sum_{s_1 \in \Sigma \setminus \{\sigma_0\}} ... \sum_{s_i \in \Sigma \setminus \{\sigma_0\}} ... {\cal P}_{1,\ell-1} \sum_{s_\ell \in \Sigma \setminus \{\sigma_0\}} p_\Sigma(S = s_\ell) \\
                                    & = & (1-p_0) \sum_{s_1 \in \Sigma \setminus \{\sigma_0\}} ... \sum_{s_i \in \Sigma \setminus \{\sigma_0\}} ... \sum_{s_{\ell-1} \in \Sigma \setminus \{\sigma_0\}} {\cal P}_{1,\ell-1} \\
                                    & = & (1-p_0)^2 \sum_{s_1 \in \Sigma \setminus \{\sigma_0\}} ... \sum_{s_i \in \Sigma \setminus \{\sigma_0\}} ... \sum_{s_{\ell-2} \in \Sigma \setminus \{\sigma_0\}} {\cal P}_{1,\ell-2} \\
                                    & = ... \\ 
                                    & = & (1-p_0)^\ell, 
\end{eqnarray*}
one can write
\begin{eqnarray}
I_0(\ell) = b + \frac{1}{(1-p_0)^\ell} \sum_{s_1, ..., s_\ell} {\cal P}_{1,\ell} \left(-\sum_{i=1}^\ell \ln p_\Sigma(S = s_i) \right). \label{temporary_information_content_versus_length_equation}
\end{eqnarray}
Notice that 
\begin{eqnarray}
\sum_{s_1, ..., s_\ell} {\cal P}_{1,\ell} (-\ln p_\Sigma(S = s_i) & = & \nonumber \\
\sum_{s_1, ..., s_{j-1}, s_{j+1},...,s_\ell} \left[ {\cal P}_{1,j-1} {\cal P}_{j+1,\ell} \sum_{s_j \in \Sigma \setminus \{\sigma_0\}} - p_\Sigma(S = s_j) \ln p_\Sigma(S = s_j) \right] & = & \nonumber \\
(H_\Sigma(S) + p_0 \ln p_0)\sum_{s_1, ..., s_{j-1}, s_{j+1},...,s_\ell} {\cal P}_{1,j-1} {\cal P}_{j+1,\ell} & = & \nonumber \\
(H_\Sigma(S) + p_0 \ln p_0)(1-p_0)^{\ell-1}, & & \label{character_entropy_extraction_equation}
\end{eqnarray}
where
\begin{eqnarray}
H_\Sigma(S) & = & - \sum_{s \in \Sigma} p_\Sigma(S = s) \ln p_\Sigma(S = s) \nonumber \\
            & = & - \sum_{s \in {\Sigma \setminus \{ \sigma_0 \}}} p_\Sigma(S = s) \ln p_\Sigma(S = s) - p_0 \ln p_0 \label{character_entropy_equation}
\end{eqnarray}
is the character entropy after the space-free prefix of length $\ell_0$.
Therefore, applying \eref{character_entropy_extraction_equation} to \eref{temporary_information_content_versus_length_equation} one finally obtains $I_0(\ell)  = a \ell + b$, where
\begin{eqnarray*}
a = \frac{1}{1-p_0} (H_\Sigma(S) + p_0 \ln p_0)
\end{eqnarray*}
and $b$ is defined as in \eref{intercept_equation}.
Notice that the slope $a$ is always positive because $H_\Sigma(S) \geq 0$ as any entropy and, according to 
\eref{character_entropy_equation}, $H_\Sigma(S) > p_0 \ln p_0$ provided that $\lambda > 1$ (recall that no character from $\Sigma$ has probability zero of occurring after the free-space prefix). Therefore, $I_0(\ell)$ grows linearly with $\ell$ for $\lambda > 1$.

\Tref{summary_table} summarizes the parameters of the linear relationship between $I_0(\ell)$ for our generalized random typing process and two particular cases: (a) equal letter probabilities (all characters except the space must be equally likely) \cite{Ferrer2009a} and (b) equal character probabilities (all characters including the space are equally likely) and empty words are allowed, i.e. $\ell_0 = 0$ \cite{Suzuki2004a}. Notice that (b) is a particular case of (a). Variant (a) \cite{Ferrer2009a} means that  
\begin{eqnarray*}
p_\Sigma(S = s) = 
    \left\{ 
    \begin{array}{cc} 
    \frac{1 - p_0}{\lambda - 1} & \qquad \mbox{if~} s \neq \sigma_0 \\
    p_0 & \qquad \mbox{if~} s = \sigma_0,
    \end{array}
    \right.
\end{eqnarray*}
and is defined only by three parameters: $\ell_0$, $\lambda$ and $p_0$. The random typing process defined in \cite{Miller1963} is a particular case with $\ell_0 = 0$.
In a random typing process with equal letter probabilities, the character entropy after the space-free prefix is
\begin{eqnarray*}
H_\Sigma(S) & = & (\lambda - 1) \left(- \frac{1-p_0}{\lambda -1} \ln \frac{1-p_0}{\lambda -1} \right) - p_0 \ln p_0 \\
            & = & (1- p_0) \ln \frac{\lambda -1}{1-p_0} - p_0 \ln p_0.
\end{eqnarray*}
Variant (b), the simplest random typing that has ever been presented to our knowledge, is defined with only one parameter, i.e. $\lambda$ ($\ell_0 = 0$ and $p_0 = 1/\lambda$ in that case). (b) is known as the fair die rolling experiment \cite{Suzuki2004a} (see \cite{Li1992b} for a version with $\ell_0 = 1$ and $p_0 = 1/\lambda$).

\begin{table}
\caption{\label{summary_table} Summary of the linear dependency between the self-information content as a function of word length, $I_0(\ell) = a + b$, and related quantities for three major variants of the random typing process.
$H_\Sigma(S)$ is the entropy of characters after the free-space prefix of length $\ell_0$, $p_0$ is the probability of space and $\lambda$ is the cardinality of $\Sigma$. $p_s$ is used as a shorthand for $p_\Sigma(S = s)$. }
\begin{indented}
\item[]\begin{tabular}{@{}lrrr}
\br
                    &             & Random typing & \\
\hline \hline
                    & Generalized & Equal letter                     & Equal character \\
                    &             & probabilities \cite{Ferrer2009a} & probabilities \\
                    &             &                                  & (with $\ell_0 = 0$ \cite{Suzuki2004a}) \\
\mr
$a$                 & $\frac{1}{1-p_0} (H_\Sigma(S)+ p_0 \ln p_0)$ & $\ln \frac{\lambda - 1}{1 - p_0}$ & $\ln \lambda$ \\ 
$b$                 & $\ln \frac{(1 - p_0)^{\ell_0}}{p_0}$ & $\ln \frac{(1 - p_0)^{\ell_0}}{p_0}$ & $\ln \lambda$ \\
$H_\Sigma(S)$       & $- \sum_{s \in {\Sigma \setminus \{ \sigma_0 \}}} p_s \ln p_s$ & $(1- p_0) \ln \frac{\lambda -1}{1-p_0}$ & $\ln \lambda$ \\ 
                    & $- p_0 \ln p_0$  & $- p_0 \ln p_0$                         & \\                 
$p_0$               & $p_0$ & $p_0$ & $\frac{1}{\lambda}$ \\
$p(W = w, \|w\| = \ell)$ & $\frac{p_0}{(1-p_0)^{\ell_0}} {\cal P}_{1,\ell}$ & $\frac{(1-p_0)^{(\ell-\ell_0)}p_0}{(\lambda - 1)^\ell}$ & $\frac{1}{\lambda}$ \\
$p(W = w| \|w\| = \ell)$ & $\frac{1}{(1-p_0)^{\ell}} {\cal P}_{1,\ell}$ & $\frac{1}{(\lambda - 1)^\ell}$ & $\frac{1}{(\lambda - 1)^\ell}$ \\
\mr
\end{tabular}
\end{indented}
\end{table}

\section{Conclusion}

We have shown that $I(\ell) = a \ell + b$ does not imply that speakers have made optimal choices as argued in \cite{Piantadosi2011a}. Uniform information density or related hypotheses (e.g., \cite{Jaeger2010a}) are not at all necesary to account for the linear correlation between $I(\ell)$ and $\ell$: typing at random yields the same dependency independently from context. Our main point is that a linear correlation between information content and word length may simply arise internally, from the units making a word (e.g., letters) and not necessarily from the interplay between words and their context as suggested in \cite{Piantadosi2011a}. However, future research should investigate if the parameters of the linear relationship predicted by random typing coincide with those of real texts or if a linear relationship is sufficient to account for the actual dependency between $I(\ell)$ and $\ell$ in real languages as it is suggested by the long-range correlations in texts at the level of words \cite{Montemurro2001b} or letters \cite{Ebeling1994,Moscoso2011a} and the differences between random typing and real language at the level of the distribution of word frequencies \cite{Ferrer2009a,Ferrer2009b} or word lengths \cite{Manin2008a}.

\ack 
This work was supported by the project OpenMT-2 (TIN2009-14675-C03) from the Spanish Ministy of Science and Innovation (RFC).

\section*{References}

\bibliographystyle{unsrt}

\end{document}